\documentclass[aps,pra,floatfix,superscriptaddress,twocolumn]{revtex4}
\usepackage[dvips]{graphicx}

\usepackage{longtable}
\usepackage{dcolumn}
\usepackage[dvips]{graphicx}
\usepackage{bm}
\usepackage{bbm}

\usepackage{times}
\usepackage{nicefrac}
\usepackage{amsmath}
\usepackage{amsfonts}
\usepackage{amssymb}
\usepackage{amsthm}

\newcolumntype{.}{D{x}{}{-1}}

\newcommand{\vare}{\varepsilon}
\newcommand{\bfr}{{\bm {r}}}

\newcommand{\bfmu}{{\bm {\mu}}}
\newcommand{\bfE}{{\bm {E}}}

\newcommand{\lbr}{\langle}
\newcommand{\rbr}{\rangle}

\newcommand{\SixJ}[6]{
        \left\{
        \begin{array}{ccc}
        #1  & #2  & #3 \\
        #4  & #5  & #6 \\
        \end{array}
        \right\}
        }

\newcommand{\Dmatrix}[4]{
        \left(
        \begin{array}{cc}
        #1  & #2   \\
        #3  & #4   \\
        \end{array}
        \right)
        }

\newcommand{\Vcase}[2]{
        \left(
        \begin{array}{c}
        #1   \\
        #2   \\
        \end{array}
        \right)
        }

\begin{document}

\title{Electric dipole polarizabilities of Rydberg states of alkali atoms}

\author{V.~A. Yerokhin}
\affiliation{Physikalisch-Technische Bundesanstalt, D-38116 Braunschweig, Germany}
\affiliation{Center for Advanced Studies, Peter the Great St.~Petersburg Polytechnic University,
195251 St.~Petersburg, Russia}

\author{S. Y. Buhmann}
\affiliation{Physikalisches Institut, Albert-Ludwigs Univesit\"at Freiburg, D-79104 Freiburg,
Germany} \affiliation{Freiburg Institute for Advanced Studies, Albert-Ludwigs Universit\"at
Freiburg, D-79104 Freiburg, Germany}

\author{S. Fritzsche}
\affiliation{Helmholtz-Institut Jena, D-07743 Jena, Germany}
\affiliation{Theoretisch-Physikalisches Institut, Friedrich-Schiller-Universit\"at Jena, D-07743
Jena, Germany}

\author{A.~Surzhykov}
\affiliation{Physikalisch-Technische Bundesanstalt, D-38116 Braunschweig, Germany}
\affiliation{Technische Univesit\"at Braunschweig, D-38106 Braunschweig, Germany}

\begin{abstract}

Calculations of the static electric-dipole scalar and tensor polarizabilities are presented for
two alkali atoms, Rb and Cs, for the $nS$, $nP_{\nicefrac{1}{2},\,\nicefrac{3}{2}}$, and
$nD_{\nicefrac{3}{2},\,\nicefrac{5}{2}}$ states with large principal quantum numbers up to $n =
50$. The calculations are performed within an effective one-electron approximation, based on the
Dirac-Fock Hamiltonian with a semi-empirical core-polarization potential. The obtained results
are compared with those from a simpler semi-empirical approach and with available experimental
data.

\end{abstract}

\maketitle

%
%
\section{Introduction}

The term Rydberg atom refers to an atom with one (or several) valence electron(s) in a state with a
large principal quantum number $n$. Such states are characterized by relatively long life time and
huge polarizabilities ($\sim\!n^7$), which result in very large responses to electric and magnetic
fields. Such exaggerated properties lead to strong, tunable interactions among the atoms, which
have applications in various fields of physics. One of the prominent examples is the effect known
as the Rydberg excitation blockade \cite{gaetan:09,urban:09}. In this effect, the excitation of
more than one Rydberg atom within a blockade volume is suppressed, since such excited states are
shifted out of resonance with a narrow-band excitation laser by the interaction between the Rydberg
atoms.

The blockade effect relies on the energy level shift that one Rydberg atom experiences in a close
proximity to another. Similar level shifts arise also when a Rydberg atom is brought into the
vicinity of a macroscopic body or surface. With the constantly increasing experimental abilities to
trap and manipulate atoms close to macroscopic bodies, the effects of the atom-surface interactions
of the van der Waals \cite{lennard32} and Casimir--Polder \cite{casimir:48} type have become a
subject of great interest. When compared to atoms in their ground  or lowest excited states, the
Rydberg atoms exhibit various peculiarities of the atom-surface interactions, such as effects from
the electric quadrupole and higher multipole moments \cite{crosse:10}, non-perturbative energy
shifts, and surface-induced state mixing \cite{ribeiro:15}.

Theoretical description of the dynamics of Rydberg atoms in complex environments as well as their
interactions with surfaces can be parameterized in terms of several basic atomic properties, such
as transition energies, dipole matrix elements and atomic polarizabilities. Their integral
convolution with the Green's tensor of the electric field in the macroscopic environment provides
\cite{agarwal:75:II,agarwal:75:IV} expressions
  for  physical observables (decay rates, environment-induced
energy shift, etc.). A classical example is the Casimir-Polder interaction potential between an
atom and a perfectly conducting wall,
\begin{align}
U(z) = &\ -\frac{\hbar}{(4\pi)^2\vare_0z^3}\,\int_0^{\infty}d\omega\,
  \alpha_0(i\omega)\,
  \nonumber \\ & \times
  \left[ 1 + 2\, \frac{\omega z}{c} + 2 \biggl( \frac{\omega z}{c}\biggr)^2 \right]
    e^{-2\omega z/c}\,,
\end{align}
which requires the detailed knowledge of the dynamic dipole
polarizability of the atom at imaginary energies $\alpha_0(i\omega)$.

In order to describe the interaction of the Rydberg atoms with a complex macroscopic environment,
first of all, we require a robust numerical approach for calculating energy levels, dipole
transition matrix elements and atomic polarizabilities. In the present paper, we develop such a
numerical approach and apply it for computations of static electric polarizabilities, for which
numerous experimental and theoretical results are available in the literature. A comparison of the
results obtained by different theoretical approaches and the experimental data helps us to
establish the level of accuracy of our treatment.

In the present work we consider the alkali atoms, with a single highly excited valence electron
beyond a closed-shell core. Their spectrum resembles that of the hydrogen atom and can be well
described by effective one-electron approximations. The approach used in this work is based on the
Dirac-Fock Hamiltonian with a semi-empirical core polarization potential. By using a finite basis
set representation of the functional space, we obtain the spectrum of the eigenvalues and
eigenfunctions the Hamiltonian, which allows us to compute various atomic properties, in
particular, polarizabilities. By comparing the results obtained by this method with those from a
simpler semi-empirical approach (the so-called Coulomb approximation), we get an idea about the
uncertainty of our treatment.

The remaining paper is organized as follows. In Sec.~\ref{sec:polarizabilities} we give the outline
of the theory of the atomic polarizabilities. The Dirac-Fock--Core-Polarization approach is
discussed in Sec.~\ref{sec:DFCP}. In Sec.~\ref{sec:Coulomb_approximation} we describe the
semi-empirical Coulomb approximation approach to the evaluation of the atomic polarizabilities. Our
results are presented and discussed in Sec.~\ref{sec:results}. The paper ends with a short
conclusion in Sec.~\ref{sec:conclusion}.

%
%
\section{Atomic polarizabilities}
\label{sec:polarizabilities}

\subsection{General theory}

Electric polarizabilities most naturally appear when considering energy shifts of atomic levels
induced by the interaction with a classical external electric field (Stark effect). In the
nonrelativistic theory, the interaction  with a static electric field $\bfE$ is given by the
operator
\begin{align} H = -\bfmu \cdot \bfE\,,
\end{align}
where $\bfmu = -e\sum_i \bfr_i$ is the electric dipole operator, $e$ is the electric charge and
$\bfr_i$ is the position vector of the $i$th electron. Due to symmetry arguments, the first-order
expectation value of $H$ on any atomic state vanishes but the higher-order perturbation
contributions survive. We are presently interested in the second-order Stark effect, caused by the
effective interaction of the form
\begin{align}\label{eq02}
H_{\rm 2Stark} = \bfmu \cdot \bfE \, \frac{1}{E_0-H_0}\, \bfmu \cdot \bfE\,,
\end{align}
where $H_0$ is the Hamiltonian of the atom in the absence of the electric field and $E_0$ denotes
its eigenvalue. The operator (\ref{eq02}) can be conveniently represented as a sum of the scalar
and the tensor parts,
\begin{align}\label{eq03}
H_{\rm 2Stark} &\ = H_0 + H_2
 \nonumber \\
&\ = \sum_{K = 0,2}\sum_{Q=-K}^K (-1)^Q\,\left\{ \bfmu\frac{1}{E_0-H_0}\bfmu\right\}^K_Q\, \left\{
\bfE\bfE\right\}^K_{-Q}\,,
\end{align}
where $H_0$ and $H_2$ correspond to $K = 0$ and $K = 2$, respectively, and $\left\{ \ldots
\right\}^K_Q$ denotes the tensor product of two vectors,
$$
\left\{ \bm{L}\bm{M} \right\}^K_Q = \sum_{qq'} C_{1q,1q'}^{KQ} \bm{L}_{q}\,\bm{M}_{q'}\,.
$$

The static electric-dipole scalar and tensor polarizabilities ($\alpha_0$ and $\alpha_2$,
respectively) are defined \cite{angel:68} from the matrix elements of $H_0$ and $H_2$ between the
$M=J$ (so-called ``stretched'') states,
\begin{align}\label{eq04}
\lbr JJ | H_0 | JJ\rbr &\ = -\frac12\,\alpha_0(J)\,E^2\,,
 \\
\lbr JJ | H_2 | JJ\rbr &\ = -\frac14\,\alpha_2(J)\,\bigl( 3E_z^2 - E^2\bigr)\,.
\end{align}
The quadratic Start shift of the energy level $(J,M)$ then takes the form \cite{angel:68}
\begin{align}\label{eq04b}
\Delta E = -\frac12\, \alpha_0(J)\,E^2 -\frac14\, \alpha_2(J) \frac{3M^2-J(J+1)}{J(2J-1)} \bigl( 3E_z^2 - E^2\bigr)\,.
\end{align}

The polarizabilities can be expressed in terms of the reduced matrix elements of the operator
$\bfmu$ as follows \cite{angel:68}
\begin{align}\label{eq05}
\alpha_0(J) = -\frac{2}{3(2J+1)}\, \sum_{n} (-1)^{J_n-J}
\frac{\lbr 0J || \bfmu ||nJ_n\rbr\lbr nJ_n || \bfmu || 0J\rbr}{E_0-E_n}\,,
\end{align}
\begin{align}\label{eq06}
\alpha_2(J) &\ = (-1)^{2J+1}\,\sqrt{\frac{40 J (2J-1)}{3(J+1)(2J+1)(2J+3)}}
  \nonumber \\ & \times
\sum_n
 \SixJ{1}{1}{2}{J}{J}{J_n}\,
\frac{\lbr 0J || \bfmu ||nJ_n\rbr\lbr nJ_n || \bfmu || 0J\rbr}{E_0-E_n}\,,
\end{align}
where the sum over $n$ implies the summation over the complete spectrum of intermediate states of
the atomic Hamiltonian. It follows from the above expression that $\alpha_2$ vanishes for $J = 0$
and $1/2$.

\subsection{Effective one-electron approximation}

In the present investigation we are interested in the highly excited Rydberg states of alkali
atoms. Such systems can be effectively described within an effective one-electron model, in which
the  valence electron interacts with the nuclear charge and the core. The direct contribution from
the core polarizability is small as compared to the valence-electron polarizability (because of the
$n^7$ scaling) and can be neglected. The indirect core-polarizability contribution, however, is
sizeable (as it modifies the valence and intermediate-state electron energies and wave functions)
and accounted for by the core polarization potential in the Hamiltonian.

In an effective one-electron approximation, the electric-dipole polarizabilities of a state $v$ can
be evaluated as
\begin{align} \label{eq2}
\alpha_0(v) = \frac{2}{3(2j_v+1)}\, \sum_n  \frac{\left[C_1(\kappa_v,\kappa_n)\,R^{(1)}_{vn}\right]^2}{\vare_n-\vare_v}\,,
\end{align}
\begin{align} \label{eq4}
\alpha_2(v) &\ = \sqrt{\frac{40 j_v (2j_v-1)}{3(j_v+1)(2j_v+1)(2j_v+3)}}
\nonumber \\ \times &
\sum_n (-1)^{j_v+j_n} \SixJ{1}{1}{2}{j_v}{j_v}{j_n}\,
\frac{\left[C_1(\kappa_v,\kappa_n)\,R^{(1)}_{vn}\right]^2}{\vare_n-\vare_v}\,,
\end{align}
where the sum over $n$ runs over the complete spectrum of virtual excited states, $\vare$ is the
energy of the corresponding state, $\kappa$ denotes the Dirac angular momentum-parity quantum
number, $j$ is the total angular momentum quantum number, $j = |\kappa|-1/2$. The radial integrals
are given by
\begin{align} \label{eq3}
R^{(L)}_{vn} = \int_0^{\infty}dr\, r^{2+L}\,W_{vn}(r)\,,
\end{align}
where
\begin{align} \label{eq2a}
W_{vn}(r) = g_v(r)\,g_n(r) + f_v(r)\, f_n(r)\,,
\end{align}
where $g(r)$ and $f(r)$ are the upper and the lower radial components of the Dirac wave function,
respectively. The angular coefficients are given by
\begin{align} \label{CJ}
C_J(\kappa_a ,\kappa_b) = (a||C^{(J)}||b) = &\  (-1)^{J}
 \sqrt{2j_a+1}\,C_{j_a \nicefrac{1}{2},J\,0}^{j_b \nicefrac{1}{2}}
 \nonumber \\  &\times
    \,\Pi (l_a,l_b,J)
 \,,
\end{align}
where $C^{(J)}$ denotes the normalized spherical harmonics, the symbol $\Pi (l_1,l_2,l_3)$ is unity
if $l_1+l_2+l_3$ is even, and zero otherwise, and $l$ is the orbital angular momentum quantum
number, $l = |\kappa+1/2|-1/2$.

In the present work, we calculate the  polarizabilities $\alpha_0$ and $\alpha_2$ of highly excited
states of alkali atoms within the effective one-electron approaches, which are discussed in the
next two sections.

%
%
\section{Dirac-Fock--Core-Polarization Hamiltonian} \label{sec:DFCP}

The radial Dirac-Fock equation with a core-polarization (CP) potential (termed hereafter as the
DFCP equation) is given by
\begin{align} \label{eq20a}
 h_{\rm DFCP}\,\phi_a(r)
 = \vare_a\, \phi_a(r)\,,
\end{align}
where $h_{\rm DFCP}$ is the Hamiltonian, $\phi_a(r)$ is a two-component radial wave function
$$
\phi_a(r) = \left( {g_a(r)}\atop{f_a(r)}\right)\,,
$$
and $\vare_a$ is the energy eigenvalue. The DFCP Hamiltonian is given by
\begin{align} \label{eq20}
h_{\rm DFCP} &= \Dmatrix{m+V(r)}{ \displaystyle-\frac1r\, \frac{d}{dr}\,r+\frac{\kappa_a}{r}}
  {\displaystyle\frac1r\,\frac{d}{dr}\,r+\frac{\kappa_a}{r}}{-m+V(r)}\,.
\end{align}
The potential $V(r)$ in the above equation is
\begin{align}
V(r) = V_{\rm nucl}(r) + V_{\rm DF}(r)+V_{\rm CP}(r)\,,
\end{align}
where $V_{\rm nucl}$ is the binding nuclear potential, $V_{\rm DF}$ is the frozen-core Dirac-Fock
potential and $V_{\rm CP}$ is the core-polarization potential. The Dirac-Fock potential for the
case of the interaction with a closed shell is defined by its matrix elements ($a\not\in c$),
\begin{align} \label{eq21}
\lbr a| V_{\rm DF}|a\rbr &\ = \alpha \sum_{c \in {\rm core}} \biggl[ (2j_c+1)\,R_0(acac)
 \nonumber \\ &
 - \frac1{2j_a+1} \sum_L \left[ C_L(\kappa_a,\kappa_c)\right]^2\, R_L(acca)\biggr]\,,
\end{align}
where  index $c$ runs over all different core states, and $R_L$ are the Slater integrals
\begin{align} \label{eq22}
R_L(abcd) = \int_0^{\infty} dr_1\,dr_2\,(r_1r_2)^2\,\frac{r_<^L}{r_>^{L+1}}\,W_{ac}(r_1)\,W_{bd}(r_2)\,.
\end{align}

The semi-empirical CP potential partly accounts for the second- and higher-order interaction of the
valence electron with the core. It is given by (see, e.g., Ref.~\cite{norcross:76,mitroy:88})
\begin{align} \label{eq23}
V_{\rm CP}(r) = -\frac{\alpha_c}{2\,r^4}\,\bigl(1-e^{-r^6/\rho_{\kappa}^6}\bigr)\,,
\end{align}
where $\alpha_c$ is the static dipole polarizability of the core and $\rho_{\kappa}$ is the radial
cutoff parameter, to be adjusted empirically.

In the present work we are interested in the complete energy spectrum of $h_{\rm DFCP}$ and the
corresponding set of eigenstates. We obtain these numerically in several steps. In the first step,
we compute the core wave functions $c$ by solving the standard Dirac-Fock equation for the
(closed-shell) core. In the second step, we solve the DFCP equation with help of the finite basis
set constructed with $B$ splines, using the core wave functions obtained in the first step for the
evaluation of the matrix elements of the DF potential. The finite basis set method provides us with
(a discrete representation of) the complete spectrum of the DFCP equation. In the third step, we
fix the empirical value of the CP cutoff parameter $\rho_{\kappa}$ (one for each $\kappa$) by
matching the experimental energies in the high-$n$ region.

The solution of the Dirac equation with a finite basis constructed with $B$ splines have been first
introduced in Ref.~\cite{johnson:88}. In present work, we solve the DFCP equation with a
modification of the $B$-splines approach, namely, the Dual Kinetic Balance (DKB) method
\cite{shabaev:04:DKB}. Within this method, the two-component solutions of Eq.~(\ref{eq20a}) are
approximated by an expansion over the finite basis of $2N$ functions $u_n$,
\begin{align} \label{eq24}
\phi_a = \sum_{n=1}^{2N} c_n\,u_n = &\ \sum_{i = 1}^N c_i\, \Vcase{B_i(r)}{\displaystyle \frac1{2m}\biggl(\frac{d}{dr}+\frac{\kappa_a}{r}\biggl)\,B_i(r)}
 \nonumber \\ &
+ \sum_{i = 1}^N c_{i+N}\, \Vcase{\displaystyle
\frac1{2m}\biggl(\frac{d}{dr}-\frac{\kappa_a}{r}\biggl)\,B_i(r)}{B_i(r)}\,,
\end{align}
where $\left\{B_i(r)\right\}_{i = 1}^N$ is the set of $B$ splines \cite{deboor:78} on the interval
$r = 0 \ldots R_{\rm max}$, where $R_{\rm max}$ is the cavity radius (chosen to be sufficiently
large in order to have no influence on the calculated properties of the atom). We note that the
anzatz (\ref{eq24}) assumes that the potential in the Dirac equation is regular at $r\to 0$. This
means that it can be used for solving the Dirac equation for an extended-nucleus potential, but
{\it not} for the point-nucleus Coulomb potential.

The expansion (\ref{eq24}) and the standard action principle lead to a generalized eigenvalue
problem for the coefficients $c_k$,
\begin{align} \label{eq24a}
\left[ \lbr u_i|h_{\rm DFCP}|u_k\rbr + \lbr u_k|h_{\rm DFCP}|u_i\rbr\right] c_k = 2\,E\,\lbr u_i|u_k\rbr\,c_k\,,
\end{align}
where the summation over repeated indices is implied and $i,k = 1\ldots 2N$. The equation
(\ref{eq24a}) is solved by the standard methods of linear algebra. Using the boundary conditions
$\phi(0) = \phi(R) = 0$, we obtain a finite basis representation of the complete spectrum of the
DFCP equation.

In the present work we are interested in highly excited Rydberg states of an atom, with the
principal quantum number up to $n = 50$. It is nontrivial task to obtain an accurate representation
of such highly excited states by a finite basis set method. In the original studies
\cite{johnson:88,shabaev:04:DKB} only the first few excited states were accurately reproduced. In
the present work, we searched for a way to increase the number of bound states in the
pseudospectrum. We found that the number of bound states depends, most pronounced, on the cavity
radius $R$ and, less so, on the radial grid and the number of basis functions. In our calculations
for Rb atom, we used the cavity radius of about $R = 4000$-$5000$~a.u. and the radial grid that is
equidistant $r\sim t$ within the nucleus and polynomial $r\sim t^4$ outside of the nucleus (where
$t$ denotes the equidistant grid). With the basis set of $N = 150-250$ $B$ splines, we obtained a
pseudospectrum with typically 60-70 bound states. We checked that, for the standard frozen-core
Dirac-Fock potential, our results for the energies of the valence excited states agree very well
with those obtained by the direct solution of the Dirac-Fock equation \cite{bratzev:77}.

After reproducing the Dirac-Fock energies, we included the CP potential. For the core
polarizability $\alpha_c$ we used the calculated results from the literature \cite{johnson:83}. The
cutoff parameter $\rho_{\kappa}$ was adjusted empirically to match the experimental energies of
$nlj$ Rydberg states for high $n$. More specifically, for each angular momentum quantum number
$\kappa$, we chose the value of $\rho_{\kappa}$ that minimized the deviation of the DFCP energies
from the experimental energies (as given by Eq.~(\ref{eq10})) for the states with $n$ from 20 to
50.

After we obtained the DFCP pseudospectrum, it is relatively straightforward to compute the
$\alpha_0$ and $\alpha_2$ polarizabilities according to Eqs.~(\ref{eq2}) and (\ref{eq4}). We would
like to stress that in the DFCP approach, we take into account the complete spectrum of the DFCP
equation, both the discrete and the continuum parts. The contribution of the continuum part of the
spectrum was found to be rather small, which justifies the usage of the so-called
``sum-over-states'' method for calculating the atomic polarizabilities
\cite{mitroy:10,safronova:11}.

%
%
\section{Coulomb approximation} \label{sec:Coulomb_approximation}

\subsection{Quantum defect energies}
\label{subsubdsec:quantum_defects}

The term ``quantum defect'' was introduced nearly a century ago by Schr\"odinger
\cite{schroedinger:21} (see Ref.~\cite{rau:97} for a historical account). Since then the concept of
the quantum defects was extensively used in the atomic physics, mostly (but not only) for the
description of energy levels of Ryberg atoms.

The quantum defect approach suggests the parametrization of the energy levels of the
valence-excited Rydberg states of alkali atoms in the form similar to that of the hydrogen atom,
\begin{align} \label{eq10}
E(n\kappa) \equiv \vare(n\kappa)-m =
-\frac{\,m_r\,\alpha^2}{2\,{{n}^*}^2}\,,
\end{align}
where $m_r$ is the reduced mass and $n^*$ is the effective (fractional) principal quantum number
which replaces the true (integer) principal quantum number $n$ in the nonrelativistic hydrogen
theory. The effective principal quantum number $n^*$ is usually parameterized in terms of the
quantum defect parameters $\delta_{i} \equiv\delta_{i}(\kappa)$ as follows
\begin{equation}
   \label{eq:principal_quantum_number}
   {n}^* = n - \sum_{i=0}^{\infty} \frac{\delta_{2i}}{(n-\delta_{0})^{2i}} \, ,
\end{equation}
In practice, the expansion over $i$ is terminated after a few first terms (usually, two).

The quantum defect parameters can be calculated by {\em ab initio} methods or, alternatively,
extracted from experimental data. It should be noted that for high Rydberg states (typically, $n
> 30$), the experimental data follow the parameterization (\ref{eq10})-(\ref{eq:principal_quantum_number}) with a very high accuracy.
For the two atoms considered here, Rb and Cs, the Rydberg spectra are extensively studied
experimentally and highly accurate results for the quantum defect parameters are available. The
literature values of quantum defect expansion coefficients used in this paper for Rb and Cs are
collected in Table~\ref{tab:QD}.

\subsection{Polarizabilities}
\label{subsubsec:Coulomb_radial_integrals}

It was demonstrated long ago \cite{bates:49} that the wave functions of excited atomic states at
large distances from the nucleus can be effectively approximated by the modified Coulomb solutions
with fractional principal quantum numbers that are obtained from the quantum defect energies
(\ref{eq10}). This simple semi-empirical approach (often referred to as the Coulomb Approximation,
CA) allows one to evaluate various transition integrals, providing that their dominant contribution
originates from large radial distances. The CA method was successively applied for calculating
atomic polarizabilities of alkali atoms \cite{wijngaarden:94,wijngaarden:97} and was shown to yield
results in a remarkably good (for such a simple approximation) agreement with experimental data. In
the present work, we perform calculations of polarizabilities by the CA method and compare the
results with those obtained with the more elaborated DFCP approach, in order to get an idea about
the uncertainties of the theoretical treatment.

In the CA method, we assume the energies of the bound state of interest to be given by
Eq.~(\ref{eq10}), with the quantum defect parameters determined from the experimental data. We now
are looking for a solution of the Schr\"odinger equation that has a fractional principal quantum
number $n^*$ and is regular at $r\to \infty$. Naturally, since the corresponding energy is not the
eigenvalue of the Schr\"odinger-Coulomb Hamiltonian, such a solution will diverge at $r\to 0$; but
it will be of no importance for us since we are interested only in the large-$r$ region. The exact
solution of the Schr\"odinger equation with the Coulomb potential regular at $r\to \infty$ can be
written as
\begin{eqnarray}
\label{eq:wave_functions_Whittaker}
G_{n^* l}(r) &=& (-1)^l \, \, n^* \, \, \Gamma[n^* + l -1] \,
 \nonumber \\ && \times
   \sqrt{\frac{\Gamma[n^* - l]}{\Gamma[n^* + l + 1]}}  \, W_{n^*\!,\, l + 1/2}(2r/n^*) \, ,
\end{eqnarray}
where the $W(x)$ is the Whittaker function, $l$ is the orbital angular momenta, and $n^*$ is the
effective principal quantum number, as obtained from Eq.~(\ref{eq:principal_quantum_number}). It
can be explicitly checked that for integer values of $n^* = n$, the function $G_{nl}(r)$ coincides
with the well known nonrelativistic bound-state wave function.

In order to compute the radial integrals required for $\alpha_0$ and $\alpha_2$, we made the
replacement
\begin{equation} \label{eq25}
R^{(L)}_{ab} \to \tilde{R}^{(L)}_{ab} = \int_{r_{\rm min}}^{\infty} dr\, r^{L} \, G_{n^*_a l_a}(r) \,
   G_{n^*_b l_b}(r) \, ,
\end{equation}
where $r_{\rm min}$ is a small radial cutoff parameter.

Following \cite{bates:49}, we compute the Whittaker $W$ function by its asymptotic expansion
\begin{eqnarray}
   \label{eq:wave_functions_series}
   W_{n^*\!,\, l + 1/2}(2r/n^*) &=& {\rm e}^{-r/n^*} \,
    \left( \frac{2r}{n^*} \right)^{n^*} \, \left[ 1 + \sum\limits_{t=1}^\infty \frac{a_t}{r^k} \right] \, ,
\end{eqnarray}
with the expansion coefficients $a_t$ obtained by the recurrence relation
\begin{eqnarray}
   \label{eq:at_coefficients}
   a_t &=& a_{t-1} \, \frac{n^*}{2 t} \,
   \left[l(l+1) - (n^* - t) (n^* - t + 1) \right] \, , \\
   a_1 &=& \frac{n^*}{2} \, \left[l(l+1) - n^* (n^* - 1) \right] \, .
\end{eqnarray}
Based on our calculations we found that summation over $t$ in Eq.~(\ref{eq:wave_functions_series})
can be truncated at about $t_{\rm max} = n^* + 1$ without losses of accuracy in the evaluation of
the radial integrals.

The integration in Eq.~(\ref{eq25}) was performed numerically by employing the Gauss--Legendre
quadratures. Following Ref.~\cite{bates:49}, we defined the lower bound of the radial integration
in Eq.~(\ref{eq25}) as $r_{\rm min} = s \; n_{a} n_{b} / (n_{a} + n_{b})$ with $s \approx 1/100$.
We have checked that the final results do not depend on the particular choice of the parameter $s$.

The sum over $n$ in Eqs.~(\ref{eq2}) and (\ref{eq4}) is performed by explicitly summing over the
virtual excited bound states (the so-called ``sum-over-states'' method). The continuum part of the
spectrum yields a very small contribution to polarizabilities and was neglected. The summation over
$n$ is extended until the convergence is reached (typically, about 20-30 virtual excited states are
included).

\section{Results and discussion}
\label{sec:results}

We start our discussion by analysing the energy values delivered by the DFCP method. Our numerical
results for energy levels of Rb and Cs are presented in Table~\ref{tab:en}, in comparison with the
values obtained from the quantum defect formula (\ref{eq10}) and the experimental quantum defect
parameters from Table~\ref{tab:QD}. We observe that the DFCP method reproduces the experimental
energies very well, provided that the principal quantum number $n$  is large enough, typically $n
\ge 20$. For smaller $n$'s, the deviation of the DFCP energies from the experimental values
gradually increases as $n$ decreases. We therefore expect that the accuracy of the DFCP results for
the polarizabilities will not depend on $n$ for $n \ge 20$  and will gradually deteriorate as $n$
is decreased from $n = 20$ downwards.

In order to estimate the uncertainty of our theoretical description of polarizabilities, we compare
the results obtained by two different methods, the DFCP and the CA ones. We expect that both
methods should become essentially equivalent in the high-$n$ limit; for smaller $n$'s, however,
some deviations are expected. The difference between the results will give us an idea about the
error of the treatment.

The comparison is presented in Table~\ref{tab:DFCP_Coulomb}. We observe that for $n\ge 15$, both
methods give results equivalent at a 1\% level for all the states considered. For the lower-$n$
states, the situation is somewhat different for Rb and Cs. For Rb, the agreement between the two
methods is very good for the $S$ and $P$ states, whereas for $D$ states there are deviations on a
few \% level. For Cs, we observe larger deviations than for Rb, which however disappear in the
high-$n$ limit.

We now compare our DFCP values of polarizabilities with  previous experimental and theoretical
results available in the literature. The comparison for $\alpha_0$ and $\alpha_2$ in Rb and Cs
 is presented in Tables~\ref{tab:a0:Rb}-\ref{tab:a2:Cs}. The complete tabulation of our DFCP
results is given in Tables~I-IV of Supplementary Material. Results are reported for the states with
the principal quantum number $n$ in the region $n = 8\,$--$\,50$ for Rb and $n = 9\,$--$\,50$ for
Cs.

We observe that for the $nS$ states, there is a very good agreement between different calculations
and also between theoretical and experimental values. In particular, we note excellent agreement
with high-precision experimental results by Walls {\em et al.} \cite{walls:01} for the $9S$ and
$10S$ states of Rb and those by Wijngaarden {\em et al.} \cite{wijngaarden:94:exp} for the $11S$,
$12S$, and $13S$ states of Cs. Based on this comparison and on the fact that the DFCP approach
works better with the increase of $n$, we estimate that the accuracy of our DFCP results for the
$nS$ states should be better than 1\% for all $n$'s presented in the tables.

For the $P$ states, there is no experimental data available. Based on the comparison presented in
Table~\ref{tab:DFCP_Coulomb}, we estimate that the accuracy of our results should be on the 1\%
level for $n\ge 15$. For the $D$ states, we observe that deviations from the experimental data are
larger than for the $S$ and $P$ states, and that they decrease more slowly with increase of $n$. We
estimate that the accuracy of our results for the $D$ states should be on the 1\% level for $n\ge
20$ and on the 2\% level for $n \approx 15$. We note some discrepancies with the CA results by
Wijngaarden \cite{wijngaarden:94} for the 10$D$ and 11$D$ states of Rb (including the overall sign
in the case of $\alpha_0$ and a factor-of-about-two difference for $\alpha_2$), which probably are
due to numerical instabilities for high $n$'s in Ref.~\cite{wijngaarden:94}.

\begin{figure*}
\centerline{
\resizebox{0.95\textwidth}{!}{%
  \includegraphics{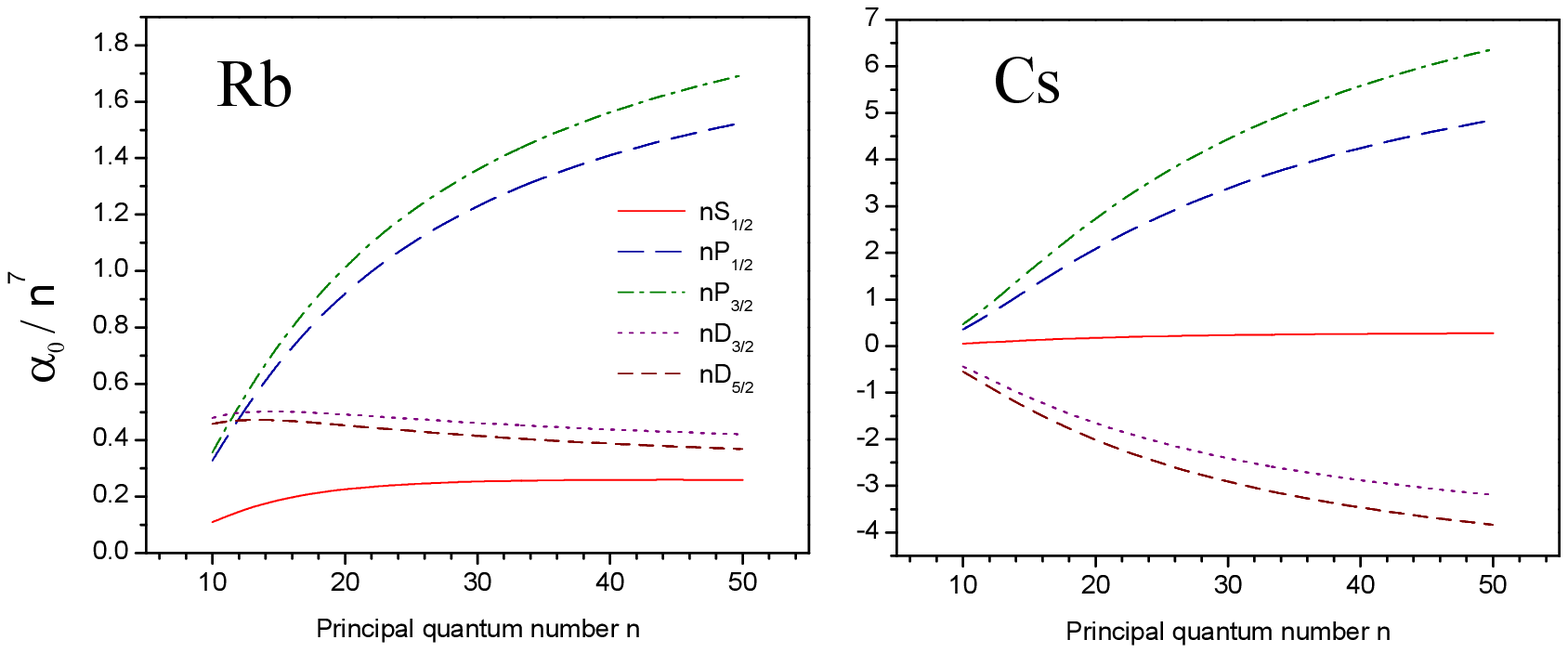}
}}
 \caption{(Color online) Electric dipole polarizabilities $\alpha_0$ scaled by the prefactor of $n^{-7}$,
as a function of the principal quantum
number $n$,  for Rb (left graph) and Cs (right graph).
 \label{fig:alpha0}}
 \end{figure*}

\begin{figure*}
\centerline{
\resizebox{0.95\textwidth}{!}{%
  \includegraphics{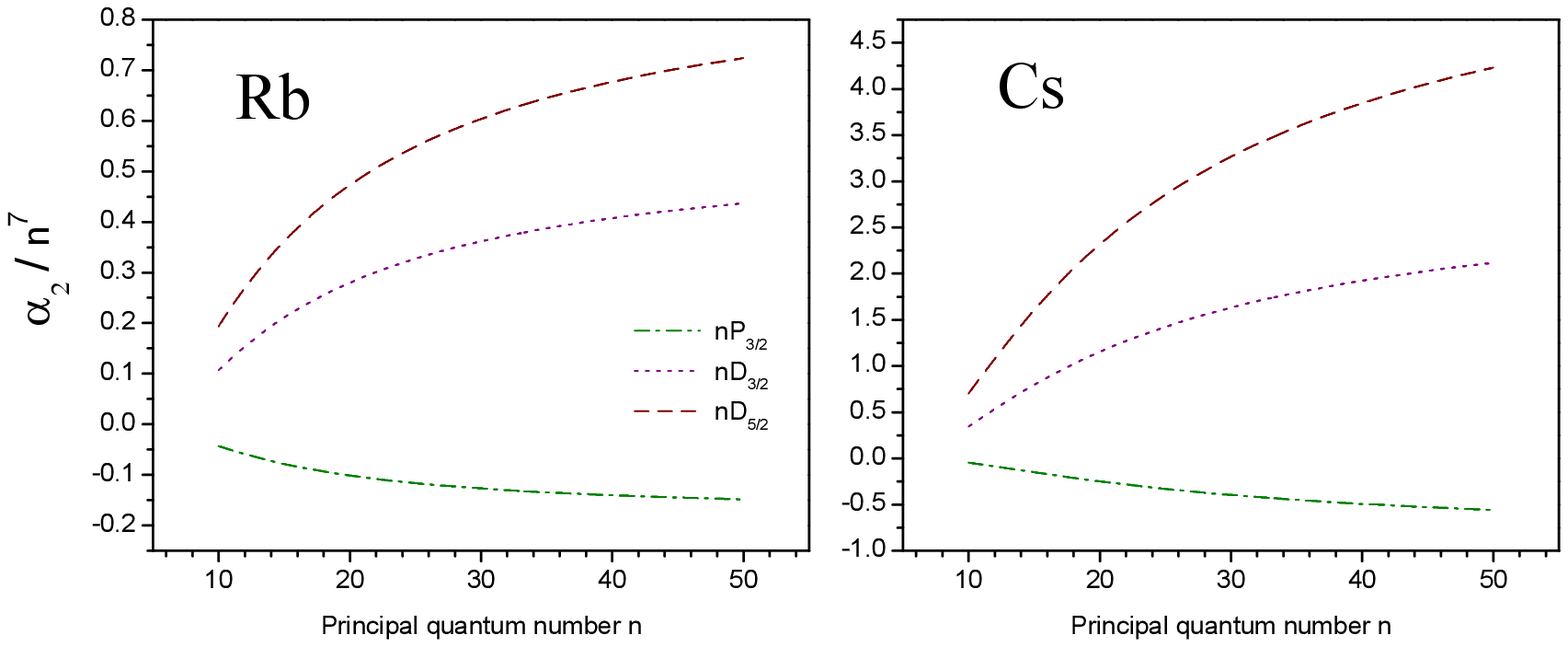}
}}
 \caption{(Color online) Electric tensor polarizabilities $\alpha_2$  scaled by the prefactor of $n^{-7}$,
as a function of the principal quantum
number $n$,  for Rb (left graph) and Cs (right graph).
 \label{fig:alpha2}}
\end{figure*}

Finally, we analyse the $n$ dependence of the electric polarizabilities. In Figs.~\ref{fig:alpha0}
and \ref{fig:alpha2} we plot our results for the $\alpha_0$ and $\alpha_2$ polarizabilities, scaled
by their leading $n$ dependence factor, $n^{-7}$. We find that for the $S$ states, the
polarizability demonstrates the asymptotic large-$n$ behaviour already at $n \approx 20$, whereas
for the $P$ and $D$ states the asymptotic behaviour is generally not reached in the range of $n \le
50$ considered in the present work.

It might be now interesting to address the question to which extent the relativistic treatment is
necessary in describing highly excited Rydberg states. In the literature, the behaviour of the
high-$n$ electrons is often considered to be non-relativistic, or even quasi-classical. From
Table~\ref{tab:en} we can deduce that the relativistic effect of the fine-structure splitting of
$nP_J$, $nD_J$, and even $nF_J$ (in the case of Cs) energy levels is rather significant on the
level of the calculational accuracy. Theoretical treatment of the atomic polarizabilities is known
to be very sensitive to the (rather small) energy difference of the reference state and the nearest
excited states of the opposite parity and, therefore, requires energy levels calculated with
inclusion of relativistic effects. Contrary to that, the transition matrix elements appearing in
the expressions for $\alpha_0$ and $\alpha_2$ are essentially non-relativistic for high $n$. This
is confirmed by good agreement observed between the DFCP and CA methods (we recall that in the CA
method, the radial integrals are calculated nonrelativistically, whereas the DFCP approach is fully
relativistic).

\section{Conslusion}
\label{sec:conclusion}

In this paper we have presented our calculations of the static electric-dipole scalar and tensor
polarizabilities of highly excited $nS$, $nP_j$, and $nD_j$ states of Rb and Cs. The calculations
are based on the Dirac-Fock Hamiltonian with a semi-empirical core-polarization potential. This
approach provides us with a complete spectrum of the energies and wavefunctions of the effective
one-particle Hamiltonian and allows us to compute various atomic properties, in particular, atomic
polarizabilities. By comparison with the results obtained by different methods and with the
experimental data, we estimate the accuracy of the obtained polarizability values to be on a \%
level for sufficiently high values of the principal quantum number $n$. In our future
investigations we plan to employ this method for computing dynamic atomic polarizabilites and
transition matrix elements necessary for the theoretical description of the interaction of Rydberg
atoms with a macroscopic environment.

\section*{Acknowledgement}

V.A.Y. acknowledges support by the Russian Federation program for organizing and carrying out
scientific investigations and by RFBR (grant No. 16-02-00538). S.Y.B. gratefully acknowledges
support by the German Research Foundation (grants BU 1803/3-1 and and GRK 2079/1) and the Freiburg
Institute for Advanced Studies.

\begin{table*}
\caption{Experimental quantum defect  parameters for Rb and Cs. \label{tab:QD}}
\begin{center}
\begin{ruledtabular}
\begin{tabular}{l.......}
  &   \multicolumn{1}{c}{$n^2S_{1/2}$}
                                & \multicolumn{1}{c}{$n^2P_{1/2}$}
                                                & \multicolumn{1}{c}{$n^2P_{3/2}$}
                                                                     & \multicolumn{1}{c}{$n^2D_{3/2}$}
                                                                                           & \multicolumn{1}{c}{$n^2D_{5/2}$}
                                                                                           & \multicolumn{1}{c}{$n^2F_{5/2}$}
                                                                                           & \multicolumn{1}{c}{$n^2F_{7/2}$}
                                                                                           \\
\hline\\[-0.2cm]
\multicolumn{8}{c}{\bf Rb}\\[0.1cm]
Ref.  &   \multicolumn{1}{c}{\cite{mack:11,li:03}}
                                & \multicolumn{1}{c}{\cite{li:03}}
                                                & \multicolumn{1}{c}{\cite{li:03}}
                                                                    & \multicolumn{1}{c}{\cite{mack:11,li:03}}
                                                                                           & \multicolumn{1}{c}{\cite{mack:11,li:03}}
                                & \multicolumn{1}{c}{\cite{han:06}}
                                & \multicolumn{1}{c}{\cite{han:06}}
                                                                                           \\[0.1cm]
$\delta_0$ & 3.131\,x180\,6(10) & 2.654\,x884\,9(10) & 2.641\,x673\,7(10) &  1.348\,x093(2)  & 1.346\,x464(2) & 0.016\,x 519\,2(9) &  0.016\,x543\,7(7)
                                                                                           \\
$\delta_2$ & 0.178\,x6\,(6)       & 0.290\,x0\,(6)       & 0.295\,x0\,(7)       & -0.604\,x2\,(13)       &-0.595\,x0\,(11)    &-0.085\,x(9)          & -0.086\,x(7)
\\                                                                                           \\
\hline\\[-0.2cm]
\multicolumn{8}{c}{\bf Cs}\\[0.1cm]
Ref.  &   \multicolumn{1}{c}{\cite{deiglmayr:16}}
                                & \multicolumn{1}{c}{\cite{deiglmayr:16}}
                                                & \multicolumn{1}{c}{\cite{deiglmayr:16}}
                                                         & \multicolumn{1}{c}{\cite{lorenzen:83}}
                                                                    & \multicolumn{1}{c}{\cite{deiglmayr:16}}
                                & \multicolumn{1}{c}{\cite{weber:87}}
                                                                                           \\[0.1cm]
$\delta_0$ & 4.049\,x353\,2(4) & 3.591\,x587\,1(3) & 3.559\,x067\,6(3) &  2.475\,x45\,(2) &  2.466\,x314\,4(6)  &  0.033\,x414(1)
                                                                                           \\
$\delta_2$ & 0.239\,x1\,(5)     & 0.362\,x73\,(16) & 0.374\,x69\,(14)  &  0.009\,x9\,(40) & 0.013\,x81\,(15)   & -0.198\,x674
                                                                                           \\
$\delta_4$ &  0.06\,(x10)        &                  &                   &-0.433\,x24       & -0.392\,x(12)       &  0.289\,x53
\\
$\delta_6$ & 11x(7)              &                  &                   &-0.965\,x55      & -1.9(3)x            &-0.260\,x1
\\
$\delta_8$ & -209x(150)          &                  &                   &-16.946\,x4
\\
\end{tabular}
\end{ruledtabular}
\end{center}
\end{table*}

\begin{table*}
\caption{Energies of the valence-excited states of Rb and Cs, for the infinitely heavy nucleus, in a.u. For each $n$,
the upper line displays the calculated DFCP energies, whereas the lower line presents the
experimental energies as obtained from Eq.~(\ref{eq10}) with the quantum defect parameteres
taken from Table~\ref{tab:QD}. The parameters of the DFCP
potential are \cite{johnson:83} $\alpha_c({\rm Rb}) = 9.076~a_0^3$, $\alpha_c({\rm Cs}) = 15.81~a_0^3$
and $\rho_{\kappa}$ as specified in the table.
\label{tab:en}}
\begin{center}
\begin{ruledtabular}
\begin{tabular}{l.......}
  &   \multicolumn{1}{c}{$n^2S_{1/2}$}
                                & \multicolumn{1}{c}{$n^2P_{1/2}$}
                                                & \multicolumn{1}{c}{$n^2P_{3/2}$}
                                                                     & \multicolumn{1}{c}{$n^2D_{3/2}$}
                                                                                           & \multicolumn{1}{c}{$n^2D_{5/2}$}
                                                                                           & \multicolumn{1}{c}{$n^2F_{5/2}$}
                                                                                           & \multicolumn{1}{c}{$n^2F_{7/2}$} \\
\hline\\[-0.2cm]
\multicolumn{8}{c}{\bf Rb}\\[0.2cm]
$\rho_{\kappa}$ & 2.4x33 &                 2.3x58         &     2.3x54    & 2.7x99          &   2.8x17  &       2.8x17   &       2.8x17
\\[3pt]
$n=20$ &  0.001\,x757\,247 &   0.001\,x662\,127 &   0.001\,x659\,599 &  0.001\,x436\,956 &  0.001\,x436\,707 &  0.001\,x252\,040 & 0.001\,x252\,045
\\
       &  0.001\,x757\,248 &   0.001\,x662\,126 &   0.001\,x659\,600 &  0.001\,x436\,953 &  0.001\,x436\,706 &  0.001\,x252\,041 & 0.001\,x252\,044
\\[3pt]
$n=30$ &  0.000\,x692\,597 &   0.000\,x668\,687 &   0.000\,x668\,042 &  0.000\,x609\,027 &  0.000\,x608\,958 &  0.000\,x556\,165 & 0.000\,x556\,166
\\
       &  0.000\,x692\,597 &   0.000\,x668\,687 &   0.000\,x668\,042 &  0.000\,x609\,033 &  0.000\,x608\,964 &  0.000\,x556\,164 & 0.000\,x556\,165
\\[3pt]
$n=40$ &  0.000\,x367\,837 &   0.000\,x358\,515 &   0.000\,x358\,262 &  0.000\,x334\,669 &  0.000\,x334\,640 &  0.000\,x312\,758 & 0.000\,x312\,758
\\
       &  0.000\,x367\,836 &   0.000\,x358\,515 &   0.000\,x358\,261 &  0.000\,x334\,672 &  0.000\,x334\,644 &  0.000\,x312\,757 & 0.000\,x312\,758
\\[3pt]
$n=50$ &  0.000\,x227\,616 &   0.000\,x223\,059 &   0.000\,x222\,934 &  0.000\,x211\,233 &  0.000\,x211\,219 &  0.000\,x200\,131 & 0.000\,x200\,131
\\
       &  0.000\,x227\,616 &   0.000\,x223\,060 &   0.000\,x222\,936 &  0.000\,x211\,235 &  0.000\,x211\,221 &  0.000\,x200\,132 & 0.000\,x200\,132
\\
\hline\\[-0.2cm]
\multicolumn{8}{c}{\bf Cs}\\[0.2cm]
$\rho_{\kappa}$ & 2.7x82 &           2.6x66              & 2.6x77&        3.1x42 &       3.1x81  &      2.9x4         &      2.9x4
\\[3pt]
$n=20$ &   0.001\,x965\,468   & 0.001\,x857\,419  & 0.001\,x850\,080 & 0.001\,x628\,123 &  0.001\,x626\,428 & 0.001\,x254\,125   & 0.001\,x254\,163
\\
       &   0.001\,x965\,462   & 0.001\,x857\,412  & 0.001\,x850\,079 & 0.001\,x628\,088 &  0.001\,x626\,393 & 0.001\,x254\,125
\\[3pt]
$n=30$ &   0.000\,x742\,480   & 0.000\,x716\,972  & 0.000\,x715\,209 & 0.000\,x659\,978 &  0.000\,x659\,539 & 0.000\,x556\,787   & 0.000\,x556\,798
\\
       &   0.000\,x742\,481   & 0.000\,x716\,973  & 0.000\,x715\,211 & 0.000\,x659\,979 &  0.000\,x659\,541 & 0.000\,x556\,787
\\[3pt]
$n=40$ &   0.000\,x386\,866   & 0.000\,x377\,201  & 0.000\,x376\,528 & 0.000\,x355\,089 &  0.000\,x354\,916 & 0.000\,x313\,021   & 0.000\,x313\,025
\\
       &   0.000\,x386\,866   & 0.000\,x377\,201  & 0.000\,x376\,528 & 0.000\,x355\,091 &  0.000\,x354\,918 & 0.000\,x313\,021
\\[3pt]
$n=50$ &   0.000\,x236\,804   & 0.000\,x232\,156  & 0.000\,x231\,831 & 0.000\,x221\,377 &  0.000\,x221\,291 & 0.000\,x200\,266   & 0.000\,x200\,268
\\
       &   0.000\,x236\,804   & 0.000\,x232\,156  & 0.000\,x231\,831 & 0.000\,x221\,378 &  0.000\,x221\,293 & 0.000\,x200\,267
\\
\end{tabular}
\end{ruledtabular}
\end{center}
\end{table*}

%
%
\begin{table*}
	\caption{Comparison of results obtained by two different methods, DFCP and CA, for the static dipole polarizabilities $\alpha_0$
of neutral Rb and Cs atoms, in $a_0^3$ ($a_0 \approx 0.052\,918$~nm is the Bohr radius). For each principal quantum number $n$,
predictions of the DFCP method are given in the upper line and those of the
CA method, in the lower line. $X[b]$ means $X\times 10^b$. \label{tab:DFCP_Coulomb}}
	\begin{center}
		\begin{ruledtabular}
			\begin{tabular}{llllll}
				$n$ &   \multicolumn{1}{c}{$n^2S$}      &   \multicolumn{1}{c}{$n^2P_{1/2}$}   &     \multicolumn{1}{c}{$n^2P_{3/2}$}
				&     \multicolumn{1}{c}{$n^2D_{3/2}$} &     \multicolumn{1}{c}{$n^2D_{5/2}$}\\
				\hline\\[-0.2cm]
\multicolumn{6}{c}{\bf Rb} \\[0.2cm]
				8 &   0.133$\,$[6]       &  0.363$\,$[6]       &  0.395$\,$[6]       & 0.909$\,$[6]       &  0.877$\,$[6]        \\
				  &   0.132$\,$[6]       &  0.364$\,$[6]       &  0.395$\,$[6]       & 0.937$\,$[6]       &  0.905$\,$[7]        \\[0.1cm]
				10 &   0.109$\,$[7]      &  0.328$\,$[7]       &  0.357$\,$[7]       & 0.479$\,$[7]       &  0.458$\,$[7]       \\
			       &   0.109$\,$[7]      &  0.327$\,$[7]       &  0.356$\,$[7]       & 0.490$\,$[7]       &  0.469$\,$[7]       \\[0.1cm]
				12 &   0.528$\,$[7]      &   0.171$\,$[8]      &  0.187$\,$[8]       & 0.178$\,$[8]       &  0.169$\,$[8]       \\
				   &   0.527$\,$[7]      &   0.171$\,$[8]      &  0.186$\,$[8]       & 0.181$\,$[8]       &  0.172$\,$[8]       \\[0.1cm]
				15 &   0.321$\,$[8]      &  0.115$\,$[9]       &  0.126$\,$[9]      &  0.858$\,$[8]       &  0.803$\,$[8]       \\
			   	   &   0.321$\,$[8]      &  0.115$\,$[9]       &  0.126$\,$[9]      &  0.867$\,$[8]       &  0.812$\,$[8]       \\[0.1cm]
				20 &   0.289$\,$[9]      &  0.118$\,$[10]      &  0.130$\,$[10]     & 0.629$\,$[9]       &  0.580$\,$[9]       \\
				   &   0.289$\,$[9]      &  0.118$\,$[10]      &  0.130$\,$[10]     & 0.631$\,$[9]       &  0.582$\,$[9]       \\[0.1cm]
				30 &   0.555$\,$[10]     &  0.269$\,$[11]      &  0.297$\,$[11]      & 0.101$\,$[11]     &  0.909$\,$[10]      \\
				   &   0.554$\,$[10]     &  0.269$\,$[11]      &  0.297$\,$[11]      & 0.101$\,$[11]     &  0.905$\,$[10]      \\[0.1cm]
\hline\\[-0.2cm]
\multicolumn{6}{c}{\bf Cs}\\[0.2cm]
				9 &  0.154$\,$[6]       &  0.105$\,$[7]       &  0.136$\,$[7]       & $-$0.146$\,$[7]       & $-$0.185$\,$[7]       \\
				  &  0.152$\,$[6]       &  0.102$\,$[7]       &  0.131$\,$[7]       & $-$0.140$\,$[7]       & $-$0.178$\,$[7]       \\[0.1cm]
				10 &  0.477$\,$[6]       &  0.356$\,$[7]       &  0.463$\,$[7]       & $-$0.436$\,$[7]       & $-$0.548$\,$[7]      \\
				   &  0.474$\,$[6]       &  0.350$\,$[7]       &  0.452$\,$[7]       & $-$0.423$\,$[7]       & $-$0.531$\,$[7]      \\[0.1cm]
				12 &  0.286$\,$[7]       &  0.246$\,$[8]       &  0.321$\,$[8]       & $-$0.254$\,$[8]       & $-$0.315$\,$[8]      \\
			  	   &  0.286$\,$[7]       &  0.243$\,$[8]       &  0.316$\,$[8]       & $-$0.250$\,$[8]       & $-$0.310$\,$[8]      \\[0.1cm]
				15 &  0.211$\,$[8]       &  0.210$\,$[9]       &  0.275$\,$[9]       & $-$0.188$\,$[9]       & $-$0.231$\,$[9]      \\
				   &  0.211$\,$[8]       &  0.209$\,$[9]       &  0.273$\,$[9]       & $-$0.187$\,$[9]       & $-$0.230$\,$[9]      \\[0.1cm]
				20 &  0.226$\,$[9]       &  0.267$\,$[10]      &  0.350$\,$[10]      & $-$0.212$\,$[10]      & $-$0.258$\,$[10]     \\
				   &  0.226$\,$[9]       &  0.266$\,$[10]      &  0.349$\,$[10]      & $-$0.211$\,$[10]      & $-$0.257$\,$[10]     \\[0.1cm]
				30 &  0.511$\,$[10]      &  0.740$\,$[11]      &  0.972$\,$[11]      & $-$0.526$\,$[11]      & $-$0.636$\,$[11]     \\
				   &  0.511$\,$[10]      &  0.740$\,$[11]      &  0.971$\,$[11]      & $-$0.527$\,$[11]      & $-$0.636$\,$[11]     \\
			\end{tabular}
		\end{ruledtabular}
	\end{center}
\end{table*}
%
%

\begin{table*}
\caption{Static electric-dipole scalar polarizabilities $\alpha_0$ (in $a_0^3$) in Rb, a comparison
with the previous experimental and theoretical results. The complete tabulation of
the DFCP results is available in Table~I of Supplementary Material.
  The notations are as follows: $X[b]$ means $X\times
  10^b$; $X(a)[b]$ means $X\times
  10^b$ with uncertainty $a$ in the last digit of $X$; ``(th)''  and ``(exp)'' refer to the theoretical and
  the experimental literature results, correspondingly. \label{tab:a0:Rb}}
\begin{center}
\begin{ruledtabular}
\begin{tabular}{lllllll}
   $n$ &   \multicolumn{1}{c}{$n^2S$}      &   \multicolumn{1}{c}{$n^2P_{1/2}$}   &     \multicolumn{1}{c}{$n^2P_{3/2}$}
   &     \multicolumn{1}{c}{$n^2D_{3/2}$} &     \multicolumn{1}{c}{$n^2D_{5/2}$} & Ref.\\
   \hline\\
  8 &   0.133$\,$[6]       &  0.363$\,$[6]       &  0.395$\,$[6]       & 0.909$\,$[6]       &  0.877$\,$[6]       \\
    &   0.132$\,$[6]      &   0.360$\,$[6]     &    0.391$\,$[6]    &    0.936$\,$[6]        & 0.904$\,$[7]     &   \cite{wijngaarden:97} (th)\\
    &   0.133(1)$\,$[6]    &                 &                 &                &           &   \cite{safronova:11} (th) \\
    &                        &                &                 &        0.927(1)$\,$[6]    &  0.8949(6)$\,$[6]     &  \cite{walls:01} (exp)\\[0.1cm]
  9 &   0.417$\,$[6]       &  0.120$\,$[7]       &  0.130$\,$[7]       & 0.221$\,$[7]       &  0.212$\,$[7]       \\
    &   0.416$\,$[6]      &   0.119$\,$[7]     &    0.129$\,$[7]    &    0.226$\,$[7]       &  0.217$\,$[7]    &   \cite{wijngaarden:97} (th) \\
    &   0.417(2)$\,$[6]    &                 &                 &                &           &   \cite{safronova:11} (th) \\
    &   0.4170(4)$\,$[6]    &                &                 &                &               &  \cite{walls:01} (exp)\\[0.1cm]
 10 &   0.109$\,$[7]       &  0.328$\,$[7]       &  0.357$\,$[7]       & 0.479$\,$[7]       &  0.458$\,$[7]       \\
    &   0.110$\,$[7]      &   0.326$\,$[7]     &    0.355$\,$[7]    & $\!\!\!\!\!-$0.485$\,$[7]    &  $\!\!\!\!\!-$0.514$\,$[7]     &   \cite{wijngaarden:97} (th)\\
    &   0.1094(6)$\,$[7]    &                 &                 &                &           &   \cite{safronova:11} (th) \\
    &   0.10953(6)$\,$[7]    &                &                 &                &               &  \cite{walls:01} (exp)\\[0.1cm]
 11 &   0.252$\,$[7]       &  0.787$\,$[7]       &  0.859$\,$[7]       & 0.956$\,$[7]       &  0.909$\,$[7]       \\
    &   0.251$\,$[7]      &   0.782$\,$[7]     &    0.854$\,$[7]    &    0.935$\,$[7]    &  $\!\!\!\!\!-$0.999$\,$[7]     &   \cite{wijngaarden:97} (th)\\[0.1cm]
 12 &   0.528$\,$[7]       &  0.171$\,$[8]       &  0.187$\,$[8]       & 0.178$\,$[8]       &  0.169$\,$[8]       \\
    &   0.526$\,$[7]      &   0.137$\,$[8]       &  0.151$\,$[8]       &                &               &   \cite{wijngaarden:97} (th)\\[0.1cm]
 13 &   0.102$\,$[8]       &  0.343$\,$[8]       &  0.375$\,$[8]       & 0.314$\,$[8]       &  0.296$\,$[8]       \\
    &             &       0.346$\,$[8]    &     0.370$\,$[8]       &                &               &   \cite{wijngaarden:97} (th)\\
    &            &                 &                    &    0.314(1)$\,$[8]     & 0.283(2)$\,$[8]     &  \cite{sullivan:85,sullivan:86} (exp) \\[0.1cm]
 15 &   0.321$\,$[8]       &  0.115$\,$[9]       &  0.126$\,$[9]      &  0.858$\,$[8]       &  0.803$\,$[8]       \\
    &   0.319(2)$\,$[8]      &     &                         &   0.860(8)$\,$[8]     & 0.796(4)$\,$[8]     &  \cite{sullivan:85,sullivan:86} (exp) \\[0.1cm]
 20 &   0.289$\,$[9]       &  0.118$\,$[10]      &  0.130$\,$[10]      & 0.629$\,$[9]       &  0.580$\,$[9]       \\
    &   0.2905(12)$\,$[9]      &     &                         & 0.643(12)$\,$[9]     &0.583(8)$\,$[9]     &  \cite{sullivan:85,sullivan:86} (exp) \\[0.1cm]
 25 &   0.149$\,$[10]      &  0.670$\,$[10]      &  0.740$\,$[10]      & 0.291$\,$[10]      &  0.264$\,$[10]      \\
    &   0.151(2)$\,$[10]      &     &                        &   0.297(12)$\,$[10]    &0.265(4)$\,$[10]     &  \cite{sullivan:85,sullivan:86} (exp) \\[0.1cm]
 30 &   0.555$\,$[10]      &  0.269$\,$[11]      &  0.297$\,$[11]      & 0.101$\,$[11]      &  0.909$\,$[10]      \\
    &   0.559(6)$\,$[10]      &     &                        &   0.104(4)$\,$[11]    & 0.936(8)$\,$[10]     &  \cite{sullivan:85,sullivan:86} (exp) \\[0.1cm]
 35 &   0.166$\,$[11]      &  0.856$\,$[11]      &  0.948$\,$[11]      & 0.289$\,$[11]      &  0.258$\,$[11]      \\
    &   0.169(1)$\,$[11]      &     &             &              0.297(8)$\,$[11]    & 0.253(8)$\,$[11]     &  \cite{sullivan:85,sullivan:86} (exp) \\[0.1cm]
 40 &   0.425$\,$[11]      &  0.231$\,$[12]      &  0.256$\,$[12]      & 0.718$\,$[11]      &  0.636$\,$[11]      \\
    &   0.425(8)$\,$[11]      &     &                        &   0.74(3)$\,$[11]     & 0.67(2)$\,$[11]     &  \cite{sullivan:85,sullivan:86} (exp) \\[0.1cm]
 45 &   0.972$\,$[11]      &  0.551$\,$[12]      &  0.611$\,$[12]      & 0.160$\,$[12]      &  0.141$\,$[12]      \\
    &   1.00(4)$\,$[11]      &     &                        &    0.169(8)$\,$[12]    & 0.153(8)$\,$[12]     &  \cite{sullivan:85,sullivan:86} (exp) \\[0.1cm]
 50 &   0.203$\,$[12]      &  0.119$\,$[13]      &  0.132$\,$[13]      & 0.329$\,$[12]      &  0.288$\,$[12]      \\
    &   0.203(1)$\,$[12]      &     &                       &    0.341(12)$\,$[12]    &0.289(16)$\,$[12]     &  \cite{sullivan:85,sullivan:86} (exp) \\[0.1cm]
\end{tabular}
\end{ruledtabular}
\end{center}
\end{table*}

\begin{table*}
\caption{Static electric-dipole tensor polarizabilities $\alpha_2$  (in $a_0^5$) in Rb, a comparison
with the previous experimental and theoretical results. The complete tabulation of
the DFCP results is available in Table~II of Supplementary Material.
Notations are as in Table~\ref{tab:a0:Rb}. \label{tab:a2:Rb}}
\begin{center}
\begin{ruledtabular}
\begin{tabular}{lllll}
   $n$ &     \multicolumn{1}{c}{$n^2P_{3/2}$}
   &     \multicolumn{1}{c}{$n^2D_{3/2}$} &     \multicolumn{1}{c}{$n^2D_{5/2}$} & Ref.\\
   \hline\\
%
  8 & $-$0.513$\,$[5]       &  0.113$\,$[6]       &  0.223$\,$[6]       \\
    & $-$0.508$\,$[5]       &  0.105$\,$[6]       &  0.211$\,$[6]      &   \cite{wijngaarden:97} (th)\\
    &                       &  0.109(6)$\,$[6]    &  0.229(12)$\,$[6]  & \cite{hogervorst:75} (exp)\\
    &                 &        0.1067(4)$\,$[6]   &  0.2086(4)$\,$[6]     &  \cite{walls:01} (exp)\\[0.1cm]
  9 & $-$0.161$\,$[6]       &  0.389$\,$[6]       &  0.723$\,$[6]       \\
    & $-$0.160$\,$[5]       &  0.373$\,$[6]       &  0.701$\,$[6]      &   \cite{wijngaarden:97} (th)\\[0.1cm]
 10 & $-$0.427$\,$[6]       &  0.107$\,$[7]       &  0.194$\,$[7]       \\
    & $-$0.424$\,$[6]       &  0.299$\,$[7]       &  0.470$\,$[7]      &   \cite{wijngaarden:97} (th)  \\[0.1cm]
 11 & $-$0.996$\,$[6]       &  0.256$\,$[7]       &  0.455$\,$[7]       \\
    & $-$0.974$\,$[6]       &  0.633$\,$[7]       &  0.998$\,$[7]      &   \cite{wijngaarden:97} (th)  \\[0.1cm]
 13 & $-$0.414$\,$[7]       &  0.110$\,$[8]       &  0.191$\,$[8]       \\
    & $-$0.366$\,$[7]       &  0.108(2)$\,$[8]    &  0.189(4)$\,$[8]    &  \cite{sullivan:86} (exp) \\[0.1cm]
 15 & $-$0.134$\,$[8]       &  0.361$\,$[8]       &  0.620$\,$[8]       \\
    &                 &  0.354(8)$\,$[8]    &  0.607(8)$\,$[8]    &  \cite{sullivan:86} (exp) \\[0.1cm]
 20 & $-$0.130$\,$[9]       &  0.359$\,$[9]       &  0.606$\,$[9]       \\
    &                 &  0.358(12)$\,$[9]    & 0.599(20)$\,$[9]   &  \cite{sullivan:86} (exp)  \\[0.1cm]
 25 & $-$0.710$\,$[9]       &  0.200$\,$[10]      &  0.335$\,$[10]      \\
    &                 &  0.201(8)$\,$[10]    & 0.326(4)$\,$[10]   &  \cite{sullivan:86} (exp)  \\[0.1cm]
 30 & $-$0.277$\,$[10]      &  0.791$\,$[10]      &  0.132$\,$[11]      \\
    &                 &  0.784(20)$\,$[10]    &0.129(4)$\,$[11]   &  \cite{sullivan:86} (exp)  \\[0.1cm]
 35 & $-$0.865$\,$[10]      &  0.249$\,$[11]      &  0.415$\,$[11]      \\
    &                 &  0.249(8)$\,$[11]    & 0.418(8)$\,$[11]   &  \cite{sullivan:86} (exp)  \\[0.1cm]
 40 & $-$0.230$\,$[11]      &  0.668$\,$[11]      &  0.111$\,$[12]      \\
    &                 &  0.64(3)$\,$[11]    &  0.11(4)$\,$[12]    &  \cite{sullivan:86} (exp) \\[0.1cm]
 45 & $-$0.541$\,$[11]      &  0.158$\,$[12]      &  0.263$\,$[12]      \\
    &                 &  0.157(12)$\,$[12]    &0.257(12)$\,$[12]  &  \cite{sullivan:86} (exp)   \\[0.1cm]
 50 & $-$0.116$\,$[12]      &  0.341$\,$[12]      &  0.566$\,$[12]      \\
    &                 &  0.329(12)$\,$[12]    &0.539(20)$\,$[12] &  \cite{sullivan:86} (exp)
\end{tabular}
\end{ruledtabular}
\end{center}
\end{table*}

\begin{table*}
\caption{Static electric-dipole scalar polarizabilities $\alpha_0$ (in $a_0^3$) in Cs, a comparison
with the previous experimental and theoretical results. The complete tabulation of
the DFCP results is available in Table~III of Supplementary Material. Notations are as in Table~\ref{tab:a0:Rb}.  \label{tab:a0:Cs}}
\begin{center}
\begin{ruledtabular}
\begin{tabular}{lllllll}
   $n$ &   \multicolumn{1}{c}{$n^2S$}      &   \multicolumn{1}{c}{$n^2P_{1/2}$}   &     \multicolumn{1}{c}{$n^2P_{3/2}$}
   &     \multicolumn{1}{c}{$n^2D_{3/2}$} &     \multicolumn{1}{c}{$n^2D_{5/2}$} & Ref.\\
   \hline\\
%
  9 &  0.154$\,$[6]       &  0.105$\,$[7]       &  0.136$\,$[7]       & $-$0.146$\,$[7]       & $-$0.185$\,$[7]       \\
    &  0.153$\,$[6]       &  0.102$\,$[7]       &  0.131$\,$[7]       & $-$0.140$\,$[7]       & $-$0.177$\,$[7]       &  \cite{wijngaarden:94} (th) \\
    &                 &                 &                 & $-$0.14(1)$\,$[7]     & $-$0.20(1)$\,$[7]     &     \cite{fredriksson:77} (exp) \\[0.1cm]
 10 &  0.477$\,$[6]       &  0.356$\,$[7]       &  0.463$\,$[7]       & $-$0.436$\,$[7]       & $-$0.548$\,$[7]       \\
    &  0.475$\,$[6]       &  0.349$\,$[7]       &  0.451$\,$[7]       & $-$0.422$\,$[7]       & $-$0.530$\,$[7]       &   \cite{wijngaarden:94} (th) \\
    &                             &                 &                 & $-$0.46(7)$\,$[7]     & $-$0.54(5)$\,$[7]     &     \cite{fredriksson:77} (exp) \\
    &                             &                 &                 & $-$0.4185(4)$\,$[7]   & $-$0.5303(8)$\,$[7]     &     \cite{xia:97} (exp) \\
    &  0.478(1)$\,$[6] &&&&&\cite{wijngaarden:94:exp} (exp) \\[0.1cm]
 11 &  0.125$\,$[7]       &  0.100$\,$[8]       &  0.131$\,$[8]       & $-$0.111$\,$[8]       & $-$0.139$\,$[8]       \\
    &  0.124$\,$[7]       &  0.099$\,$[8]       &  0.128$\,$[8]       & $-$0.109$\,$[8]       & $-$0.136$\,$[8]       &    \cite{wijngaarden:94} (th) \\
    &                             &                 &                 & $-$0.1083(1)$\,$[8]   & $-$0.1358(2)$\,$[8]     &     \cite{xia:97} (exp) \\
    &  0.1245(1)$\,$[7] &&&&&\cite{wijngaarden:94:exp} (exp) \\[0.1cm]
 12 &  0.286$\,$[7]       &  0.246$\,$[8]       &  0.321$\,$[8]       & $-$0.254$\,$[8]       & $-$0.315$\,$[8]       \\
    &  0.284$\,$[7]       &  0.244$\,$[8]       &  0.316$\,$[8]       & $-$0.251$\,$[8]       & $-$0.311$\,$[8]       &    \cite{wijngaarden:94} (th) \\
    &                             &                 &                 & $-$0.2484(2)$\,$[8]   & $-$0.3078(6)$\,$[8]     &     \cite{xia:97} (exp) \\
    &  0.2867(2)$\,$[7] &&&&&\cite{wijngaarden:94:exp} (exp) \\[0.1cm]
 13 &  0.598$\,$[7]       &  0.543$\,$[8]       &  0.709$\,$[8]       & $-$0.530$\,$[8]       & $-$0.655$\,$[8]       \\
    &  0.590$\,$[7]       &  0.540$\,$[8]       &  0.703$\,$[8]       & $-$0.522$\,$[8]       & $-$0.647$\,$[8]       &    \cite{wijngaarden:94} (th) \\
    &                             &                 &                 & $-$0.5198(7)$\,$[8]   & $-$0.643(1)$\,$[8]     &     \cite{xia:97} (exp) \\
    &  0.5993(5)$\,$[7] &&&&&\cite{wijngaarden:94:exp} (exp) \\[0.1cm]
 14 &  0.116$\,$[8]       &  0.110$\,$[9]       &  0.144$\,$[9]       & $-$0.103$\,$[9]       & $-$0.127$\,$[9]       \\
    &  0.114$\,$[8]       &  0.110$\,$[9]       &  0.143$\,$[8]       &      &&\cite{wijngaarden:94} (th) \\    [0.1cm]
 15 &  0.211$\,$[8]       &  0.210$\,$[9]       &  0.275$\,$[9]       & $-$0.188$\,$[9]       & $-$0.231$\,$[9]       \\
    &  0.206$\,$[8]       &&&&&       \cite{wijngaarden:94} (th) \\[0.1cm]
 16 &  0.365$\,$[8]       &  0.378$\,$[9]       &  0.496$\,$[9]       & $-$0.328$\,$[9]       & $-$0.403$\,$[9]       \\
    &  0.354$\,$[8]       &&&&&       \cite{wijngaarden:94} (th) \\[0.1cm]
 17 &  0.605$\,$[8]       &  0.651$\,$[9]       &  0.853$\,$[9]       & $-$0.550$\,$[9]       & $-$0.672$\,$[9]       \\
    &  0.577$\,$[8]       &&&&&       \cite{wijngaarden:94} (th) \\[0.1cm]
 39 &  0.355$\,$[11]      &  0.573$\,$[12]      &  0.753$\,$[12]      & $-$0.390$\,$[12]      & $-$0.469$\,$[12]      \\
    &                 &                 &                 & $-$0.45(2)$\,$[12]    & $-$0.49(2)$\,$[12] & \cite{zhao:11} (exp) \\[0.1cm]
 50 &  0.215$\,$[12]      &  0.379$\,$[13]      &  0.498$\,$[13]      & $-$0.249$\,$[13]      & $-$0.300$\,$[13]      \\
    &                 &                 &                 & $-$0.206(6)$\,$[13]    & $-$0.28(1)$\,$[13] & \cite{zhao:11} (exp)
\end{tabular}
\end{ruledtabular}
\end{center}
\end{table*}

\begin{table*}
\caption{Static electric-dipole tensor polarizabilities $\alpha_2$ (in $a_0^5$) in Cs, a comparison
with the previous experimental and theoretical results. The complete tabulation of
the DFCP results is available in Table~IV of Supplementary Material. Notations are as in Table~\ref{tab:a0:Rb}. \label{tab:a2:Cs}}
\begin{center}
\begin{ruledtabular}
\begin{tabular}{lllll}
   $n$ &   \multicolumn{1}{c}{$n^2P_{3/2}$}
   &     \multicolumn{1}{c}{$n^2D_{3/2}$} &     \multicolumn{1}{c}{$n^2D_{5/2}$} & Ref.\\
   \hline\\
%
%
  9 & $-$0.138$\,$[6]       &  0.121$\,$[7]       &  0.245$\,$[7]       \\
    & $-$0.134$\,$[6]       &  0.119$\,$[7]       &  0.238$\,$[7]    &   \cite{wijngaarden:94} (th) \\[0.1cm]
 10 & $-$0.460$\,$[6]       &  0.347$\,$[7]       &  0.701$\,$[7]       \\
    & $-$0.449$\,$[6]       &  0.341$\,$[7]       &  0.685$\,$[7]    &   \cite{wijngaarden:94} (th) \\
    &                       &  0.3401(4)$\,$[7]   &  0.682(2)$\,$[7]     &     \cite{xia:97} (exp) \\[0.1cm]
 11 & $-$0.127$\,$[7]       &  0.860$\,$[7]       &  0.174$\,$[8]       \\
    & $-$0.125$\,$[7]       &  0.852$\,$[7]       &  0.171$\,$[8]    &   \cite{wijngaarden:94} (th) \\
    &                       &  0.847(1)$\,$[7]    &  0.1705(5)$\,$[8]     &     \cite{xia:97} (exp) \\[0.1cm]
 12 & $-$0.308$\,$[7]       &  0.192$\,$[8]       &  0.387$\,$[8]       \\
    & $-$0.305$\,$[7]       &  0.191$\,$[8]       &  0.383$\,$[8]    &   \cite{wijngaarden:94} (th) \\[0.1cm]
%
 13 & $-$0.673$\,$[7]       &  0.393$\,$[8]       &  0.792$\,$[8]       \\
    & $-$0.670$\,$[7]       &  0.389$\,$[8]       &  0.785$\,$[8]    &   \cite{wijngaarden:94} (th) \\
    &                       &  0.3866(7)$\,$[8]    & 0.780(2)$\,$[8]     &     \cite{xia:97} (exp) \\[0.1cm]
 14 & $-$0.136$\,$[8]       &  0.753$\,$[8]       &  0.152$\,$[9]       \\
    & $-$0.136$\,$[8]       &&&       \cite{wijngaarden:94} (th) \\
    &                 &                 &  0.149(8)$\,$[9] &     \cite{fredriksson:77} (exp) \\[0.1cm]
 15 & $-$0.257$\,$[8]       &  0.136$\,$[9]       &  0.274$\,$[9]       \\
    &                 &                 &  0.28(2)$\,$[9] &     \cite{fredriksson:77} (exp) \\[0.1cm]
 16 & $-$0.460$\,$[8]       &  0.235$\,$[9]       &  0.474$\,$[9]       \\
    &                 &                 &  0.48(2)$\,$[9] &     \cite{fredriksson:77} (exp) \\[0.1cm]
 17 & $-$0.786$\,$[8]       &  0.391$\,$[9]       &  0.786$\,$[9]       \\
    &                 &                 &  0.80(4)$\,$[9] &     \cite{fredriksson:77} (exp) \\[0.1cm]
 18 & $-$0.129$\,$[9]       &  0.627$\,$[9]       &  0.126$\,$[10]      \\
    &                 &                 &  0.130(6)$\,$[10] &     \cite{fredriksson:77} (exp) \\[0.1cm]
 30 & $-$0.865$\,$[10]      &  0.357$\,$[11]      &  0.714$\,$[11]      \\
    &                 &  0.361(8)$\,$[11]   &  0.70(1)$\,$[11] & \cite{lei:95} (exp) \\[0.1cm]
 35 & $-$0.288$\,$[11]      &  0.115$\,$[12]      &  0.231$\,$[12]      \\
    &                 &  0.125(5)$\,$[12]   &  0.235(4)$\,$[12] & \cite{lei:95} (exp) \\[0.1cm]
 39 & $-$0.664$\,$[11]      &  0.261$\,$[12]      &  0.521$\,$[12]      \\
    &                 &  0.30(2)$\,$[12]    &  0.56(1)$\,$[12] & \cite{zhao:11} (exp) \\[0.1cm]
\end{tabular}
\end{ruledtabular}
\end{center}
\end{table*}

\end{document}